\documentclass[aps,prb,amsmath,amssymb,twocolumn,showpacs]{revtex4-1}
\usepackage{bm}
\usepackage{graphicx}
\usepackage{epstopdf}
\usepackage{amsmath}
\usepackage[unicode=true,colorlinks=true]{hyperref}
\begin{document}
\title{Weak localization in low-symmetry quantum wells}
\author{F. V. Porubaev}
\author{L. E. Golub}
\email{golub@coherent.ioffe.ru}
\affiliation{Ioffe Physical-Technical Institute of the Russian Academy of Sciences, 194021 St.~Petersburg, Russia}%

\begin{abstract}
Theory of weak localization is developed for electrons in semiconductor quantum wells grown along [110] and [111] crystallographic axes. Anomalous conductivity correction caused by weak localization is calculated for symmetrically doped quantum wells. The theory is valid for both ballistic and diffusion regimes of weak localization in the whole range of classically weak magnetic fields.
We demonstrate that in the presence of bulk inversion asymmetry the magnetoresistance is negative: The linear in the electron momentum spin-orbit interaction has no effect on the conductivity while the cubic in momentum coupling suppresses weak localization without a change of the correction sign.
Random positions of impurities in the doping layers in symmetrically doped quantum wells produce electric fields which result in position-dependent Rashba coupling.
This random spin-orbit interaction leads to spin relaxation which changes the sign of the anomalous magnetoconductivity. The obtained expressions allow determination of electron spin relaxation times in (110) and (111) quantum wells from transport measurements.
\end{abstract}

\maketitle

\section{Introduction}
After intensive study of spin-dependent phenomena in traditional nanostructures in the last decade,~\cite{Dyakonov_book} the attention of semiconductor spintronics community is shifting now to specially designed low-symmetrical systems where more subtle spin effects can be observed. 
One of the most interesting structures are quantum wells (QWs) grown in [110] direction.
Symmetrical (110) QWs  have unusual spin properties due to absence of relaxation for spin component oriented along the growth direction. Indeed, the bulk inversion asymmetry (BIA) 
terms in the spin-orbit Hamiltonian which lead to the D'yakonov-Perel' spin relaxation have the following form contrasted with traditional QWs:~\cite{Pikus_110,DK}
\begin{equation}
\label{H_BIA}
	H_{\rm BIA}^{(110)} =  \sigma_z (\beta k_x + \hbar \Omega_3 \cos{3\varphi_{\bm k}}). 
\end{equation}
Here we introduce crystallographic axes $z \parallel [110]$, ${x \parallel [\bar{1}10]}$, $y \parallel [001]$, 
$\beta$ is the BIA spin-orbit constant for two dimensional electrons, $\varphi_{\bm k}$ is an angle between the wavevector $\bm k$ and $x$ axis, $\Omega_3$ is a cubic function of  $k$, 
and $\sigma_z$ is the Pauli matrix. 
The form of this Hamiltonian implies that the eigenstates have a definite spin projection $\pm 1/2$ onto the growth axis $z$, which results in the absence of the D'yakonov-Perel' spin relaxation mechanism for the normal spin component.
Therefore the spin relaxation times in symmetrical (110) QWs are very long up to hundreds of nanoseconds as it has been demonstrated experimentally.~\cite{Japan99,Belkov110,Oest,Korn_Glazov,Korn_Tarasenko}

Spin relaxation can be caused by structure inversion asymmetry (SIA)
which arises from asymmetrical doping, difference of the QW barrier materials, etc.
SIA leads to a splitting of the electron energy spectrum due to the Rashba spin-orbit interaction~\cite{SD_LG}
\begin{equation}
\label{H_SIA}
	H_{\rm SIA} =  \alpha (\sigma_x k_y - \sigma_y k_x), 
\end{equation}
where $\alpha$ is a constant. 
However, there is an additional source for spin relaxation even in nominally symmetric QWs. If a QW is symmetrically doped with the doping layers located at equal distances from the center of the QW,
the system is symmetric only in average. There are domains of the nonzero electric field produced by non-mirror symmetric impurity distributions in the layers. The impurity electric fields result in a random spin-orbit coupling via  the Rashba effect because $\alpha \neq 0$ in each domain. The corresponding contribution to the Hamiltonian is given by~\cite{Glazov_review}
 \begin{equation}
\label{H_SO_random}
H_{\rm rand}({\bm r}) =  -{\rm i} \sigma_x \left\{{\partial_y}, \alpha({\bm r}) \right\}  + {\rm i}\sigma_y \left\{ {\partial_x}, \alpha({\bm r}) \right\}\:,
\end{equation}
where the Rashba-term coefficient $\alpha(\bm r)$ is coordinate dependent, and the anticommutators are defined according to $\left\{ A,B \right\}=(AB+BA)/2$. This term leads to a finite relaxation rate for the spin $z$-component. 

In the other low-symmetry heterosystem, QWs grown along the [111] crystallographic direction, the BIA spin-orbit terms have the following form:~\cite{DK}
\begin{equation}
\label{H_111}
	H_{\rm BIA}^{(111)} = \tilde{\beta} (\sigma_x k_y - \sigma_y k_x) +  \hbar \Omega_3 \cos{3\varphi_{\bm k}} \sigma_z. 
\end{equation}
Here the $\bm k$-linear contribution with a constant $\tilde{\beta}$ is of the same form as the SIA term Eq.~\eqref{H_SIA}, and the $\bm k$-cubic term has the same structure as in (110) QWs. It follows from Eqs.~\eqref{H_SIA} and~\eqref{H_111} that in structure-asymmetric (111) QWs the $\bm k$-linear terms may cancel each other if the condition  $\alpha = -\tilde{\beta}$ is fulfilled,~\cite{Cartoixa} for a review of recent experiments see Ref.~\onlinecite{Touluse_Nat_Comm_111} and references therein. In this situation the spin-orbit interaction Hamiltonian is given by Eq.~\eqref{H_BIA} with $\beta=0$, and only random spin-orbit coupling Eq.~\eqref{H_SO_random} can be responsible for relaxation of the spin $z$ component.

The spin-orbit interaction can be probed in both 
optical and transport experiments, and the low-temperature resistance measurements in classically-weak magnetic fields serves as a powerful tool for its study.
It is well known that the magnetoresistance is caused by weak localization effect consisting in the interference of electron scattering paths. 
In high-mobility structures the interfering paths may consist of a few ballistic parts in contrast to large diffusive trajectories typical for systems with low mobility, the so-called diffusion regime. This fact changes drastically the theoretical approach to the weak localization problem because the conductivity depends on the
perpendicular magnetic field $B$  
in the whole range of classically-weak fields until the magnetic length is smaller than the mean free path $l$,
i.e. up to  $B>B_{tr}$, where the ``transport'' field $B_{tr}=\hbar/(2el^2)$,~\cite{GZ,DKG} (see also Refs.~\onlinecite{Kawabata,DyakonovSSC,Cassam_Chenay_Shapiro}).
The non-diffusive theory of weak localization has been developed for traditional (001) QWs with spin-orbit splitting,~\cite{LG_05,MG_LG_FTP,SST_Review} Si-based structures,~\cite{Pudalov} $p$-type QWs,~\cite{Porubaev} topological insulators and graphene.~\cite{MN_NS_ST_SSC,MN_NS_EPL,Kachor_Karlsruhe}
Exact theoretical expressions for the weak-localization induced conductivity correction allow for accurate determination of the spin-orbit 
splittings in various two-dimensional systems such as InGaAs-, GaN- and HgTe-based QWs.~\cite{Guzenko_InGaAs,Stud_InGaAs,Koga_InGaAs,HgTe_APL,Spirito_GaN,HgTe_JAP}

In the presence of BIA spin-orbit interaction~\eqref{H_BIA}, the eigenstates at any wavevector $\bm k$ have the spin projection $\pm 1/2$ onto the $z$ axis, and 
 the interference in these two subsystems occurs independently. Due to absence of spin-flip processes, the magnetoconductivity is positive in this case like in systems without spin-orbit coupling. Moreover, at $\Omega_3=0$, the energy spectrum is parabolic with the origin shifted by $\mp m\beta/\hbar^2$ in the $x$ direction in the $\bm k$-space, where $m$ is the electron effective mass. Therefore the conductivity correction is independent of the linear spin-orbit splitting constant $\beta$. However, the $\bm k$-cubic term changes the correction in each subband. Its effect on weak localization has been taken into account in Ref.~\onlinecite{Pikus_110} but in the diffusion approximation only.

In wide asymmetrical QWs SIA is dominant, and the spin splitting is given by a homogeneous Rashba constant $\alpha \gg \beta$. In this case the conductivity correction 
is described by the theory which has been developed 
in Refs.~\onlinecite{LG_05,SST_Review} and valid for QWs of any crystallographic orientation. The magnetoconductivity can be both positive or negative depending on the relation between the spin relaxation rate $\propto \alpha^2$ and the dephasing rate $1/\tau_\phi$.

In the presence of random Rashba spin-orbit interaction~\eqref{H_SO_random}, the form of magnetocoductivity strongly depends on the correlation length $l_c$ of the spin-orbit disorder. 
In the case of small correlation length, $l_c \ll l$,  electrons travel ballistically along many domains with random Rashba splitting. The spin-orbit interaction~\eqref{H_SO_random} results in small rotations of electron spins which can be effectively described by spin-flip processes.~\cite{Glazov_review} 
Effect of spin-flip processes on weak localization has been considered in the classical work Ref.~\onlinecite{HLN} but the magnetoconductivity has been calculated there only for diffusion regime which is not realized in high-mobility QWs. An attempt to develop a theory for both ballistic and diffusion regimes has been performed in Ref.~\onlinecite{Romanov} for a particular choice of the spin-flip probability form but analytical results have not been obtained. 
In the present work we calculate the weak localization induced magnetoresistance in the whole range of classically-weak magnetic fields in (110) QWs and in (111) QWs with $\alpha=-\tilde{\beta}$. 

The paper is organized as follows.
In Sec.~\ref{BIA_cubic} we investigate the effect of the $\bm k$-cubic BIA term on the conductivity correction, in Sec.~\ref{random} we study weak localization in the presence of random spin-orbit splitting, and in Sec.~\ref{BIA_linear_random} we analyze a joint action of $\bm k$-linear BIA and random SIA splittings. The results are discussed in Sec.~\ref{disc}, and Sec.~\ref{concl} concludes the paper.

\section{Weak localization in the presence of BIA spin-orbit splitting}
\label{BIA_cubic}

\begin{figure}[t]
\includegraphics[width=0.9\linewidth]{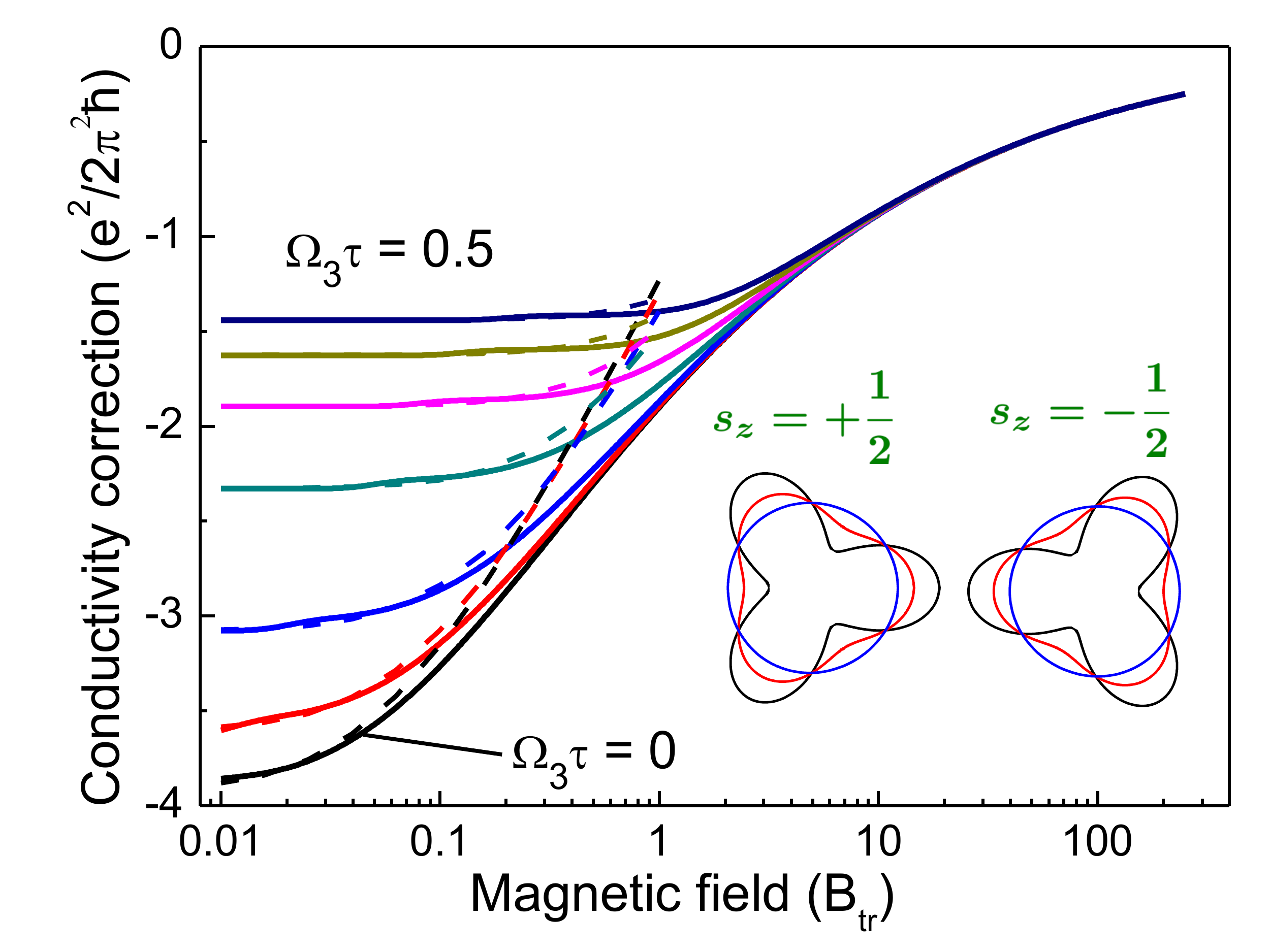}
\caption{Magnetoconductivity with account for $\bm k$-cubic BIA
spin-orbit interaction with $\Omega_3\tau=0, 0.05,0.1,0.2,0.3,0.4,0.5$ (from bottom to top) at the dephasing rate $\tau/\tau_\phi=0.01$. 
Solid lines represent the exact result, dashed lines are obtained in the diffusion approximation.
Inset: Fermi contours in two spin subsystems at $\hbar\Omega_3/E_{\rm F}$=0.01, 0.2 and 0.5.}
\label{fig_cubic_mc}
\end{figure}

As it was discussed above, the $\bm k$-linear BIA spin-orbit interaction~\eqref{H_BIA} has no effect the conductivity correction. Therefore the BIA coupling manifests itself in weak localization due to the $\bm k$-cubic splitting only. The term  $\Omega_3$ in  Eq.~\eqref{H_BIA} results in trigonal warping of the energy spectrum  opposite in two subbands, see inset to Fig.~\ref{fig_cubic_mc}, similar to other subsystems with three-fold rotation symmetry, e.g. graphene valleys.~\cite{graphene_warping}
We do not take into account the effect of $\bm k$-linear term on the $\bm k$-cubic one 
provided the splitting $2\beta k_{\rm F} \ll E_{\rm F}$, where $k_{\rm F}$ and $E_{\rm F}$ are the  Fermi wavevector and the Fermi energy, respectively.
The $\bm k$-cubic BIA term changes the interference in both spin subsystems equally, therefore we derive all expressions for the spin-up subband, and then multiply the result by two.
Assuming $\hbar\Omega_3 \ll E_{\rm F}$, we obtain
a phase factor in the electron Green functions like in the presence of other spin-orbit coupling terms:~\cite{LG_05}
\begin{equation}
\label{G}
	G^{R,A}(\bm r, \bm r') = G_0^{R,A}(\bm r, \bm r') \exp{\left({\rm i}\Omega_3\tau  \cos{3\theta}R/ l\right) }.
\end{equation}
Here $G_0^{R,A}(\bm r, \bm r')$ are the retarded and advanced Green functions at $\Omega_3=0$, $\tau$ is the momentum scattering time by a short-range potential, and $\theta$ is the angle between the vector $\bm R = \bm r-\bm r'$ and the $x$ axis. 
The key quantity for calculation of weak-localization correction to the conductivity is the so-called return probability ${P(\bm r, \bm r')= (\hbar^3 / m\tau)G^R(\bm r, \bm r')G^A(\bm r, \bm r')}$. It is given by
\begin{equation}
	P(\bm r, \bm r') = P_0(\bm r, \bm r')\exp{\left(2{\rm i}\Omega_3\tau  \cos{3\theta}R/ l\right)},
\end{equation}
where
\begin{equation}
\label{P_0}
	P_0(\bm r, \bm r') = {\exp[-R /\tilde{l}  + {\rm i}(y+y')(x'-x) / l_B^2] \over 2\pi R l}.
\end{equation}
Here ${\tilde{l}=l/(1+\tau/\tau_\phi)}$, $l_B=\sqrt{\hbar /|eB|}$ is the magnetic length, $e<0$ is the electron charge, we use the Landau gauge with the vector potential ${\bm A=(-yB,0,0)}$, and $B>0$ is assumed.

Weak localization correction to the conductivity is determined by the interference amplitude of electronic waves propagating along two time-inverted scattering paths, so-called Cooperon.
Equation for the Cooperon in the basis of Landau-level states with a double electron charge has the form~\cite{LG_05}
\begin{equation}
	C_{NN'} = \delta_{NN'} + \sum_{N_1} P_{NN_1} C_{N_1N'},
\end{equation}
where $P_{NN'}$ are the coefficients of expansion of the function $P(\bm r, \bm r')$ in this basis.
Hereafter we assume $\Omega_3\tau \ll 1$, and  hold the terms up to quadratic in $\Omega_3$. In this approximation we obtain:
\begin{align}
\label{P}
&	P_{NN'} =  \int d\bm R  {\exp(-R /\tilde{l}) \over 2\pi R l} \exp{[{\rm i}(N'-N)\theta]} f_{NN'}(R/l_B) \nonumber \\
	& \times \left[1 + 2{\rm i}\Omega_3\tau {R\over l} \cos{3\theta} - \left(\Omega_3\tau {R\over l} \right)^2 (1 + \cos{6\theta})\right] . 
\end{align}
Here $f_{NN'}(t)$ is given by~\cite{LG_05}
\begin{equation}
\label{fNN}
	f_{NN'}(t) = \left\{
	\begin{array}{c}
		 (-t)^{N'-N} {\rm e}^{-t^2/2} L_N^{N'-N}(t^2) \sqrt{N!\over N'!} \, @ N'> N,\\
		 t^{N-N'} {\rm e}^{-t^2/2} L_{N'}^{N-N'}(t^2) \sqrt{N'!\over N!} \, @ N \geq N'
	\end{array}
	 \right.
\end{equation}
 with $L_N$ and $L_N^m$ being the Laguerre polynomials.

The off-diagonal terms in the Cooperon equation appear only due to spin-orbit interaction which is small. Therefore in the second order in $\Omega_3$ we obtain:
\begin{align}
	&C_{NN} = 1 + P_{NN} C_{NN} + \sum_{N_1 \neq N} P_{NN_1} C_{N_1N}, \\
	&C_{N_1N} \approx P_{N_1N} C_{NN} \quad (N_1\neq N), \nonumber
\end{align}
and get the diagonal components in the form
\[	C_{NN} = {1 \over 1 - P_{NN} - \sum\limits_{N_1 \neq N} P_{NN_1} P_{N_1N}}.
\]
This equation demonstrates that the diagonal coefficients $P_{NN}$ should be calculated in the second order in $\Omega_3\tau$ while the off-diagonal terms $P_{NN_1}$ and $P_{N_1N}$ are needed in the first order. Therefore
only $N_1 = N\pm 3$ yields the contribution to the Cooperon.
It follows from Eq.~\eqref{P} that
\begin{align}
	P_{NN} &= P_N - (\Omega_3\tau)^2 P_N', \\
	P_{N+3,N} &= -P_{N,N+3} = {\rm i} \Omega_3\tau T_{N},
\nonumber
\end{align}
where 
\begin{equation}
\label{PN}
	    P_N = {l_B \over l} \int\limits_0^\infty dx
    \exp{\left( -x {l_B \over \tilde{l} } - {x^2 \over 2}\right)}
    L_N(x^2),
\end{equation}
\begin{equation}
	P_N' = \left( {l_B \over l} \right)^3 \int\limits_0^\infty dx
    \exp{\left( -x {l_B \over l} - {x^2 \over 2}\right)} x^2
    L_N(x^2),
\end{equation}
and
\begin{align}
	T_{N} =  \left({l_B \over l}\right)^2 {1\over\sqrt{(N+1)(N+2)(N+3)}} \\
	\times \int\limits_0^\infty dx
    \exp{\left( -x {l_B \over l} - {x^2 \over 2}\right)}
    x^4 L_{N}^3(x^2). \nonumber
\end{align}
As a result, we obtain
\begin{equation}
\label{C_cub}
	C_{NN} = {1 \over 1 - P_N + (\Omega_3\tau)^2 (P_N' - T_{N-3}^2 - T_{N}^2)}.
\end{equation}

Since $\hbar\Omega_3 \ll E_{\rm F}$, the spin-orbit interaction does not enter into
the ``vertex parts'' of the conductivity diagrams, and the weak localization induced conductivity correction is given by the same expression as in the absence of the spin-orbit splitting:~\cite{GZ}
\begin{equation}
\label{sigma}
	\sigma(B) = {e^2 \over  2\pi^2 \hbar} \left( {l \over l_B}
    \right)^2 \sum\limits_{N=0}^\infty \left( Q_N^2 {C_N+C_{N+1}\over 2} - P_N^2 C_N\right),
\end{equation}
where the spin splitting $\Omega_3$ enters into the Cooperon $C_N=2P_NC_{NN}$  (the factor 2 accounts for two spin subbands).
Here $Q_N$ is given by
\begin{equation}
\label{QN}
	Q_N= {1 \over \sqrt{N+1}} {l_B \over l}\int\limits_0^\infty dx
    \exp{\left( -x {l_B \over l} - {x^2 \over 2}\right)}
    x L_N^1(x^2).
\end{equation}

Equations~\eqref{PN}-\eqref{QN} describe the weak-localization correction to conductivity in both ballistic and diffusive regimes, i.e. valid in classically-weak fields at arbitrary values of $B/B_{tr}$.
The obtained results demonstrate that the effect of trigonal warping is not reduced to an additional dephasing as in the diffusion regime. In contrast, Eq.~\eqref{C_cub} shows that nonzero $\Omega_3$ changes the magnetoconductivity in high-mobility QWs due to magnetic field dependence of $P_N'$ and $T_N$.

At large $N$ we use the asymptotics 
\[
{x^m \over \sqrt{N^m}}L_N^m(x^2) \approx J_m(2x\sqrt{N}),
\]
and get
\begin{align}
\label{appr_low_field}
&	P_N \approx \tilde{P}_N = {1\over \sqrt{(1+\tau/\tau_\phi)^2+4N(l/l_B)^2}}, \\
&	Q^2_N \approx \tilde{P}_N^2 {1-\tilde{P}_N \over 1+\tilde{P}_N},	\quad P'_N \approx \tilde{P}_N^3 (3\tilde{P}_N^2 -1),
 \nonumber \\
&  
T_N \approx Q_N^{3}  (3/\tilde{P}_N+1) \nonumber. 
\end{align}
We use these relations in calculations of the conductivity at $N>200$.

The magnetoconductivity calculated by Eqs.~\eqref{PN}-\eqref{QN} is plotted in Fig.~\ref{fig_cubic_mc} for different strengths of the cubic spin-orbit splitting $\Omega_3$ by solid lines. The magnetoconductivity is positive at any value of the $\bm k$-cubic splitting but its absolute value decreases with increasing of $\Omega_3\tau$.

In high fields $B \gg B_{tr}$, the correction is independent of $\Omega_3$ and has the same high-field asymptotics as at zero spin splitting:~\cite{GZ} 
\begin{equation}
\label{hf}
	\sigma(B \gg B_{tr}) =-0.25 {e^2\over \hbar} \sqrt{B_{tr} \over B}.
\end{equation}

In low fields $B \sim B_{tr} \, {\tau \over \tau_\phi} \ll B_{tr}$, the difference $\sigma(B) - \sigma(0)$ is determined by $N \ll B_{tr}/B$. Therefore the diffusion approximation is valid:
\begin{equation}
\label{diff_approx}
	Q_N \approx T_{N} \approx 0, \quad
	P'_N \approx 2, \quad
	P_N \approx 1-{B\over B_{tr}}(N+1/2) - {\tau \over \tau_\phi},
\end{equation}
and we get the  expression obtained in Ref.~\onlinecite{Pikus_110}:
\begin{equation}
\label{diff_cub}
	\sigma(B) - \sigma(0) =  {e^2 \over 2 \pi^2 \hbar} 
 f_2\left({B/B_{tr} \over \tau/\tau_\phi + 2 (\Omega_3\tau)^2} \right) ,
\end{equation}
where 
$f_2(x)$ is given by 
\begin{equation}
\label{f2}
	f_2(x) = \psi(1/2+1/x) + \ln{x}
\end{equation}
with $\psi(y)$ being the digamma-function.
Dashed lines in Fig.~\ref{fig_cubic_mc} represent results of diffusion approximation~\eqref{diff_cub}. 
One can see that the diffusion approximation describes the exact result at $B \ll B_{tr}$  in the range where the conductivity weakly depends on the field.
For higher magnetic fields one should use Eqs.~\eqref{PN}-\eqref{QN} for calculation of the magnetoconductivity.

In zero field, very large $N$ are important, and we can perform integration over $N$ instead of summation in Eq.~\eqref{sigma}. This procedure yields:
\begin{equation}
\label{zero-field-cub}
	\sigma(0) = -{e^2 \over \pi^2 \hbar} \int\limits_0^{1-\tau/\tau_\phi} dP {P\over 1+P} {1\over 1- P + (\Omega_3\tau)^2 (P'-2T^2)}.
\end{equation}
Here  $P'$ and $T$ are related to $P$ by Eqs.~\eqref{appr_low_field}
which are correct in the zero-field limit too.
It is seen that the $\bm k$-cubic splitting influences the weak localization in a more complicated manner than just an additional dephasing, see Eq.~\eqref{zero-field-cub}, in contrast to the diffusion-approximation result of Ref.~\onlinecite{Pikus_110}.
At $\Omega_3=0$ the result is given by
\begin{equation}
\label{sigma0}
	\sigma(0)	= -{e^2 \over 2\pi^2 \hbar} \ln{\tau_\phi\over 2\tau}.
\end{equation}

\begin{figure}[t]
\includegraphics[width=0.9\linewidth]{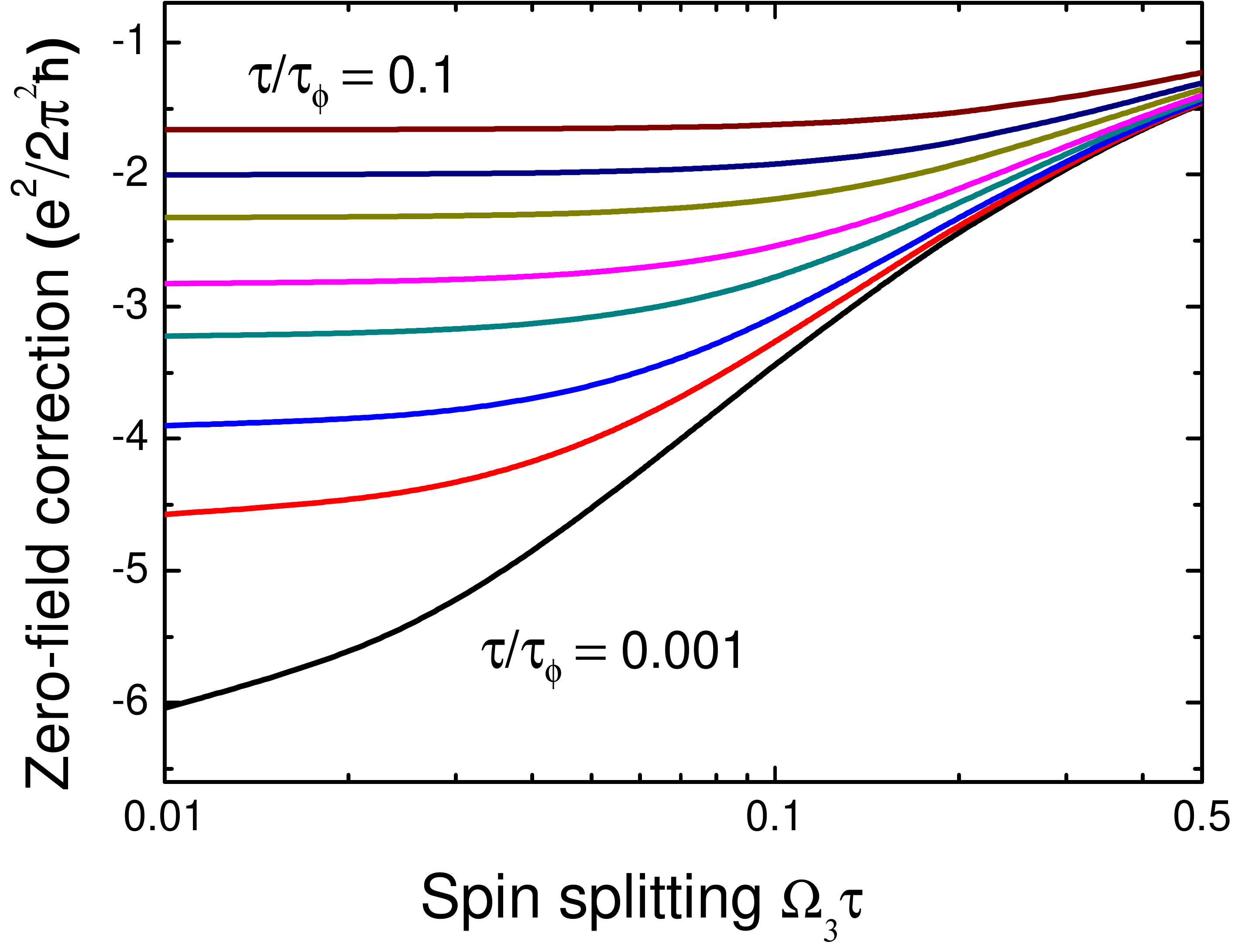}
\caption{Zero-field conductivity correction in the presence of $\bm k$-cubic term $\Omega_3$. The dephasing rate $\tau/\tau_\phi=0.001,0.005,0.01,0.02,0.03,0.05,0.07,0.1$ (from bottom to top).}
\label{fig_cubic_zero_field}
\end{figure}

The zero-field correction to the conductivity is plotted in Fig.~\ref{fig_cubic_zero_field}. One can see that the $\bm k$-cubic spin splitting destroys weak localization decreasing the conductivity correction absolute value but does not change its sign.

\section{Random spin-orbit splitting}
\label{random}

In this Section, we investigate weak localization effect in the presence of spin-orbit disorder.
At random Rashba interaction~\eqref{H_SO_random} with a position-dependent factor $\alpha(\bm r)$  zero on average,
a spin-dependent phase does not appear in the Green functions in contrast to Eq.~\eqref{G} but the spin-orbit coupling manifests itself in the
matrix element of spin-dependent scattering:~\cite{Glazov_review}
\begin{equation}
\label{V_sf}
	V_{\bm k', \bm k} = \alpha(\bm k' - \bm k) (\sigma_x K_y - \sigma_y K_x).
\end{equation}
Here $\alpha(\bm q)$ is the Fourier image of $\alpha(\bm r)$ introduced in Eq.~\eqref{H_SO_random}, and ${\bm K = (\bm k+\bm k')/2}$.

We see that the weak localization problem in the presence of the random Rashba interaction is equivalent to that at spin-flip scattering. A similar problem beyond the diffusion approximation has been considered in Ref.~\onlinecite{Romanov} where it has been shown that the Cooperon depends on three arguments: the positions of the first and last scatterers $\bm r_1$ and $\bm r_2$ as well as on the unit vector $\bm e_{i1}$ pointing from the electron position $\bm r_i$ before the first scattering event to $\bm r_1$. The Cooperon equation has the following form:
\begin{align}
\label{C}
&	C(\bm r_1, \bm r_2|\varphi) = \delta(\bm r_1 - \bm r_2) \\
&	+ \int d\bm r_3  P_0(\bm r_1,\bm r_3) 
\left({W_{\bm k' \bm k}\over W_0} - {\tau\over 2\tau_s}\right) 
C(\bm r_3, \bm r_2|\varphi').
	\nonumber
\end{align}
Here $P_0(\bm r_1,\bm r_3)$ is given by Eq.~\eqref{P_0},
the factor $\tau/(2\tau_s)$ takes into account that the total electron scattering rate is a sum of the spin-independent rate $1/\tau$ and spin-flip rate $1/2\tau_s$,
$W_0=\hbar^3/(m\tau)$ is the correlator of spin-independent scattering potential assumed to be short-range, 
$\varphi$ and $\varphi'$ are the azimuthal angles of the vectors $\bm e_{i1}$ and $\bm r_{13}$,
and 
\[W_{\bm k' \bm k} 	= W_0 + V_{-\bm k', -\bm k} \otimes V_{\bm k', \bm k}.
\]
According to Eq.~\eqref{V_sf} we have
\begin{align}
&	W_{\bm k' \bm k}  =W_0 - {\cal K}_\alpha (|\bm k' - \bm k|) 
	\Biggl[ {\sigma_x\rho_x + \sigma_y\rho_y \over 2}K^2 \nonumber \\
&	- {\sigma_x\rho_x -\sigma_y\rho_y \over 2} (K_x^2 - K_y^2)
	-(\sigma_x\rho_y + \sigma_y\rho_x) K_xK_y\Biggr].\nonumber 
\end{align}
Here $\sigma_{x,y}$ and $\rho_{x,y}$ are the Pauli matrices acting on the two first and two last spin indices, respectively, and 
\[{\cal K}_\alpha(q) = \int d {\bm r} \, \left<\alpha({\bm r}) \alpha({\bm r}')\right> \, {\rm e}^{{\rm i} {\bm q} \cdot ({\bm r} - {\bm r}')}
\] 
is the Fourier component of the correlator of the Rashba spin-splittings 
with angular brackets denoting  averaging over the spin-orbit disorder.
The relaxation rate for the spin $z$-component at low temperatures is expressed via ${\cal K}_\alpha(q)$ as follows:~\cite{Glazov_review}
\begin{equation}
\label{sf_rate}
	{1\over \tau_s} = {m \over 2\hbar^3} \left< {\cal K}_{\alpha}(| {\bm k}' - {\bm k} |) \left({\bm k}' + {\bm k}\right)^2 \right>_{\varphi_{\bm k},\varphi_{\bm k'}},
\end{equation}
where  the angular brackets denote averaging over directions of $\bm k$ and $\bm k'$ at the Fermi circle.
The correlator ${\cal K}_{\alpha}(q)=2\pi l_c^2 \left< \alpha^2 \right> \exp(-ql_c)$,~\cite{Glazov_review} which yields for the spin-flip time $\tau_s \gg \tau$  the following expression 
\begin{equation}
\label{tau_sf}
	{1\over \tau_s}= {4m\left< \alpha^2 \right> \over \hbar^3} k_{\rm F}l_c.
\end{equation}

Introducing the operator of the total angular momentum of two interfering particles ${\bm S = (\bm \sigma + \bm \rho)/2}$, we can rewrite $W_{\bm k' \bm k}$ as
\begin{align}
\label{W}
	W_{\bm k' \bm k} = W_0 + {\cal K}_\alpha (|\bm k' - \bm k|) \Biggl[
	(I -S_x^2-S_y^2)K^2 \\
	+ (S_x^2-S_y^2)(K_x^2-K_y^2) + 4 \{S_xS_y\}K_xK_y
	\Biggr]. \nonumber
\end{align}
We solve Eq.~\eqref{C} expanding the Cooperon by the Landau-level functions of an electron with a double elementary charge. It follows from  Eq.~\eqref{C} that the expansion coefficients $C_{NN'}(\varphi)$ satisfy the following equation: 
\begin{multline}
\label{C_N}
	C_{NN'}(\varphi) = \delta_{NN'} \\
	+ \sum_{N_1} \Pi_{NN_1} \Biggl[ \left(1 - {\tau\over 2\tau_s} \right) \left< {\rm e}^{{\rm i}(N_1-N)\varphi'} C_{N_1N'}(\varphi')\right>_{\varphi'} \\
	+ \left< {\rm e}^{{\rm i}(N_1-N)\varphi'} U(\varphi, \varphi') C_{N_1N'}(\varphi')\right>_{\varphi'} \Biggr],
\end{multline}
where we take into account the factor $\tau/\tau_s \ll 1$ in the first order only. Here
\[
\Pi_{NN'} =  l^{-1}\int_0^\infty dR  \exp(-R /\tilde{l})  f_{NN'}(R/l_B)
\]
with $f_{NN'}$ given by Eq.~\eqref{fNN},
and the dimensionless function $U(\varphi, \varphi')$ defined as $W_{\bm k' \bm k}/W_0=1+U(\varphi, \varphi')$
has the form
\begin{multline}
\label{U}
	U(\varphi, \varphi') = {\tau\over 2\tau_s} {{\cal K}_\alpha (\varphi - \varphi') \over \left<{\cal K}_\alpha (\theta)(1+\cos{\theta})\right>_\theta }  \\
	\times \biggl\{
(I -S_x^2-S_y^2)[1+\cos{(\varphi - \varphi')}]\\
	 + {1\over 2} (S_x^2-S_y^2)[\cos{2\varphi}+\cos{2\varphi'}+2\cos{(\varphi+\varphi')}] \\
	+ \{S_xS_y\}[\sin{2\varphi}+\sin{2\varphi'}+2\sin{(\varphi+\varphi')}]
	\biggr\}. 
\end{multline}
Here ${\cal K}_\alpha$ depends on the scattering angle ${\theta=\varphi - \varphi'}$ because, for elastic scattering at the Fermi circle, the change of the wavevector $q=2k_{\rm F}|\sin{(\theta/2)}|$.

It follows from Eq.~\eqref{C_N} that the $\varphi$-dependence of $C_{NN'}$ appears due to weak spin-orbit interaction only. Hence $C_{NN'}(\varphi) = \overline{C}_{NN'} + \delta C_{NN'}(\varphi)$, where the second term is much smaller than the first one. 
Therefore the equations for $\overline{C}_{NN'}$ and $\delta C_{NN'}(\varphi)$
have the following form in the first order in $\tau/\tau_s \ll 1$:
\begin{multline}
	\overline{C}_{NN'} = \delta_{NN'} 
	+  P_N\left(1 - {\tau\over 2\tau_s} \right)\overline{C}_{NN'} \\
	+ \sum_{N_1} \Pi_{NN_1}  \left< {\rm e}^{{\rm i}(N_1-N)\varphi'} \delta C_{N_1N'}(\varphi')\right>_{\varphi'} \\
	+ \sum_{N_1} \Pi_{NN_1}  \left< {\rm e}^{{\rm i}(N_1-N)\varphi'} U(\varphi, \varphi')\right>_{\varphi,\varphi'} \overline{C}_{N_1N'} , \nonumber \\
	\delta C_{NN'}(\varphi) = \sum_{N_2} \Pi_{NN_2} \left< {\rm e}^{{\rm i}(N_2-N)\varphi'} \delta U(\varphi, \varphi')\right>_{\varphi'} \overline{C}_{N_2N'}. 
\end{multline}
Here we  take into account that the first term in the sum in Eq.~\eqref{C_N} is independent of $\varphi$ and, hence, does not enter into equation for $\delta C_{NN'}$.
The function $\delta U(\varphi, \varphi')$ is defined as
\[\delta U(\varphi, \varphi')= U(\varphi, \varphi') - \left< U(\varphi, \varphi')\right>_{\varphi}.
\]
Substituting $\delta C_{NN'}(\varphi)$ into the equation for $\overline{C}_{NN'}$ we obtain:
\begin{multline}
\label{C_av}
	\overline{C}_{NN'}=  \delta_{NN'} 
	+  P_N\left(1 - {\tau\over 2\tau_s} \right)\overline{C}_{NN'} \\
	+ \sum_{N_1} \Pi_{NN_1} \left< {\rm e}^{{\rm i}(N_1-N)\varphi'} U(\varphi, \varphi')\right>_{\varphi,\varphi'} \overline{C}_{N_1N'} \\
	+ \sum_{N_1,N_2} \Pi_{NN_1}  \Pi_{N_1N_2} \left< {\rm e}^{{\rm i}(N_1-N)\varphi} {\rm e}^{{\rm i}(N_2-N_1)\varphi'} \delta U(\varphi, \varphi') \right>_{\varphi,\varphi'} \\
	\times \overline{C}_{N_2N'} .
\end{multline}

Since $U$ and $\delta U$ depend on the operators $S_i$, the above equation has the matrix form in the basis of four two-particle states. However, for the singlet state $S_i=0$, and we have an independent equation with
\begin{align}
&	U_0(\theta=\varphi-\varphi') = {\tau\over 2\tau_s} {{\cal K}_\alpha (\theta)(1+\cos{\theta}) \over \left<{\cal K}_\alpha (\theta)(1+\cos{\theta})\right>_\theta }, \\
&	\delta U_0(\theta) = U_0(\theta) - {\tau\over 2\tau_s}. \nonumber
\end{align}
The solution for the singlet Cooperon has the form
\begin{equation}
\label{singlet_Cooperon}
	\overline{C}_{NN'}^{(0)}= {\delta_{NN'} \over 1- P_N - {\tau \over 4\tau_s} d_N	},
\end{equation}
where
\begin{equation}
	d_N = \sum\limits_{N_1 \neq N} \Pi_{NN_1} \Pi_{N_1 N}
{\left<{\cal K}_\alpha (\theta)d_{NN_1}(\theta)\right>_\theta \over \left<{\cal K}_\alpha (\theta)(1+\cos{\theta})\right>_\theta }.
\end{equation}
Here $d_{NN_1}(\theta) = \cos{(N_1-N-1)\theta} + \cos{(N_1-N+1)\theta} + 2\cos{(N_1-N)\theta}$.
The correlator ${\cal K}_\alpha$ is independent of $\theta$ for small domain correlation length $l_c k_{\rm F}\ll 1$,~\cite{Glazov_review} therefore we get in this limit:
\begin{equation}
	d_N = - (Q_N^2+Q_{N-1}^2),
\end{equation}
where $Q_N$ is given by Eq.~\eqref{QN}.
The presence of the spin relaxation rate in the singlet Cooperon is caused by the difference of both the departure time $\tau_0$ and the transport  time $\tau_{\rm tr}$ from $\tau$ in the presence of spin-flip scattering:
\[
{\tau \over \tau_0} = 1 + {\tau \over 2\tau_s}, \qquad {\tau \over \tau_{\rm tr}} = 1 + {\tau \over 4\tau_s}, 
\qquad {\tau_{\rm tr} \over \tau_0} \approx 1 + {\tau \over 4\tau_s}.
\]
Note that  the singlet contribution depends on $\tau_s$ only in the fields $B \sim B_{tr}$ where $d_N \sim 1$.

It follows from Eq.~\eqref{U} that for the triplet Cooperon with zero $z$-projection of the momentum ($S=1,S_z=0$) we also have an independent equation with 
\begin{equation}
	U_{1,0}(\theta)= -U_0(\theta), \qquad \delta U_{1,0}(\theta) = -\delta U_0(\theta).
\end{equation}
Therefore the triplet Cooperon $\overline{C}_{NN'}^{(1,0)}$ is given by:
\begin{equation}
	\overline{C}_{NN'}^{(1,0)}=  {\delta_{NN'}\over 1- P_N + {\tau\over \tau_s}P_N + {\tau\over 4\tau_s} d_N }.
\end{equation}

For triplet Cooperons with a nonzero $z$-projection of the momentum ($S=1,S_z=\pm 1$) we have a set of coupled equations.
The matrix $U_{1,\pm 1}$ in the basis of these two states has the following form:
\begin{equation}
	U_{1,\pm 1}(\varphi,\varphi') = {\tau\over 2\tau_s} {{\cal K}_\alpha (\varphi-\varphi') \over \left<{\cal K}_\alpha (\theta)(1+\cos{\theta})\right>_\theta } 
	\left(
		\begin{array}{cc}
		0&\gamma\\
		\gamma^*&0
\end{array}
	\right),
\end{equation}
where
\[	\gamma = {\rm e}^{-{\rm i}(\varphi+\varphi')} + {1\over 2}{\rm e}^{-2{\rm i}\varphi} + {1\over 2}{\rm e}^{-2{\rm i}\varphi'}.
\]
Calculating the two average values entering into Eq.~\eqref{C_av}, we obtain the following equation for the triplet Cooperons with $S_z=\pm 1$:
\begin{align}
\label{C_1_pm}
&	\overline{C}_{NN'} = \delta_{NN'} 
	+  P_N\left(1 - {\tau\over 2\tau_s} \right)\overline{C}_{NN'} \\
&	+ {\tau\over 4\tau_s} \left[ 
	\left(
		\begin{array}{cc}
		0&b_N\\
		0&0
\end{array}
	\right) \overline{C}_{N+2,N'} 
	+ \left(
		\begin{array}{cc}
		0&0\\
		b_{N-2}&0
\end{array}
	\right) \overline{C}_{N-2,N'}
	\right]. \nonumber
\end{align}
Here 
\begin{align}
	b_N = &{\left<{\cal K}_\alpha (\theta)(1+\cos{2\theta}+2\cos{\theta})\right>_\theta \over \left<{\cal K}_\alpha (\theta)(1+\cos{\theta})\right>_\theta } \Pi_{N,N+2} \\
	&+ \sum_{N_1 \neq N} \Pi_{NN_1}\Pi_{N_1, N+2} {\left<{\cal K}_\alpha (\theta)b_{NN_1}(\theta)\right>_\theta \over \left<{\cal K}_\alpha (\theta)(1+\cos{\theta})\right>_\theta }, \nonumber 
	\end{align}
where we used the property $\Pi_{NN'}=(-1)^{N-N'}\Pi_{N'N}$ and introduced $b_{NN_1}(\theta)=\cos{(N_1-N)\theta}+\cos{(N_1-N-2)\theta}+2\cos{(N_1-N-1)\theta}$.
For $l_c k_{\rm F}\ll 1$ we obtain
\begin{equation}
	b_N = S_N(1+P_{N+2}) + 2Q_NQ_{N+1},
\end{equation}
where 
$P_N$ and $Q_N$ are given by Eqs.~\eqref{PN} and~\eqref{QN}, respectively, and

\begin{multline}
	S_N = {1 \over \sqrt{(N+1)(N+2)}} {l_B \over l} \\
	\times \int\limits_0^\infty dx
    \exp{\left( -x {l_B \over \tilde{l} } - {x^2 \over 2}\right)}
    x^2 L_N^2(x^2).
\end{multline}
From the matrix form of Eq.~\eqref{C_1_pm} is seen that the equations for (++) and ($-$+) elements are separated:
\begin{align}
\label{CCC}
	a_N \overline{C}_{NN'}^{(++)} - {\tau \over 4\tau_s} b_N \overline{C}_{N+2,N'}^{(-+)} &= \delta_{NN'}, \\
	a_N \overline{C}_{NN'}^{(-+)} - {\tau \over 4\tau_s} b_{N-2} \overline{C}_{N-2,N'}^{(++)} &=0. \nonumber
\end{align}
Taking $N=N'$ in the first equation and $N=N'+2$ in the second equation, we obtain a closed system for $\overline{C}_{N'N'}^{(++)}$ and $\overline{C}_{N'+2,N'}^{(-+)}$ and find:
\begin{equation}
	\overline{C}_{NN}^{(++)} = {a_{N+2} \over a_{N+2}a_N - \left({\tau \over 4\tau_s} b_N \right)^2},
\end{equation}
where 
\begin{equation}
	a_N=1-P_N+{\tau \over 2\tau_s}P_N.
\end{equation}
Performing analogous manipulations with equations for (+$-$) and ($--$) elements, we get:
\begin{equation}
	\overline{C}_{NN}^{(--)} = {a_{N-2} \over a_{N-2}a_N - \left({\tau \over 4\tau_s} b_{N-2} \right)^2}.
\end{equation}

The conductivity correction is given by Eq.~\eqref{sigma} with 
$C_N = P_N \left( \overline{C}_{NN}^{(++)} + \overline{C}_{NN}^{(--)} + \overline{C}_{NN}^{(1,0)} - \overline{C}_{NN}^{(0)}\right)$. Substitution yields:
\begin{align}
\label{C_sf}
		C_{N} &=  {P_N} \\
&		\times \Biggl[ {1 \over 1 - (1-{\tau \over \tau_s})P_N + {\tau \over 4\tau_s}d_N}
		- {1 \over 1 - P_N - {\tau \over 4\tau_s}d_N} \nonumber \\
&	+ {a_{N+2} \over a_{N+2}a_N - \left({\tau \over 4\tau_s} b_N \right)^2}
	+ {a_{N-2} \over a_{N-2}a_N - \left({\tau \over 4\tau_s} b_{N-2} \right)^2} \Biggr], \nonumber 
	\end{align}
where all values with negative indices should be substituted by zeros.

\begin{figure}[t]
\includegraphics[width=0.9\linewidth]{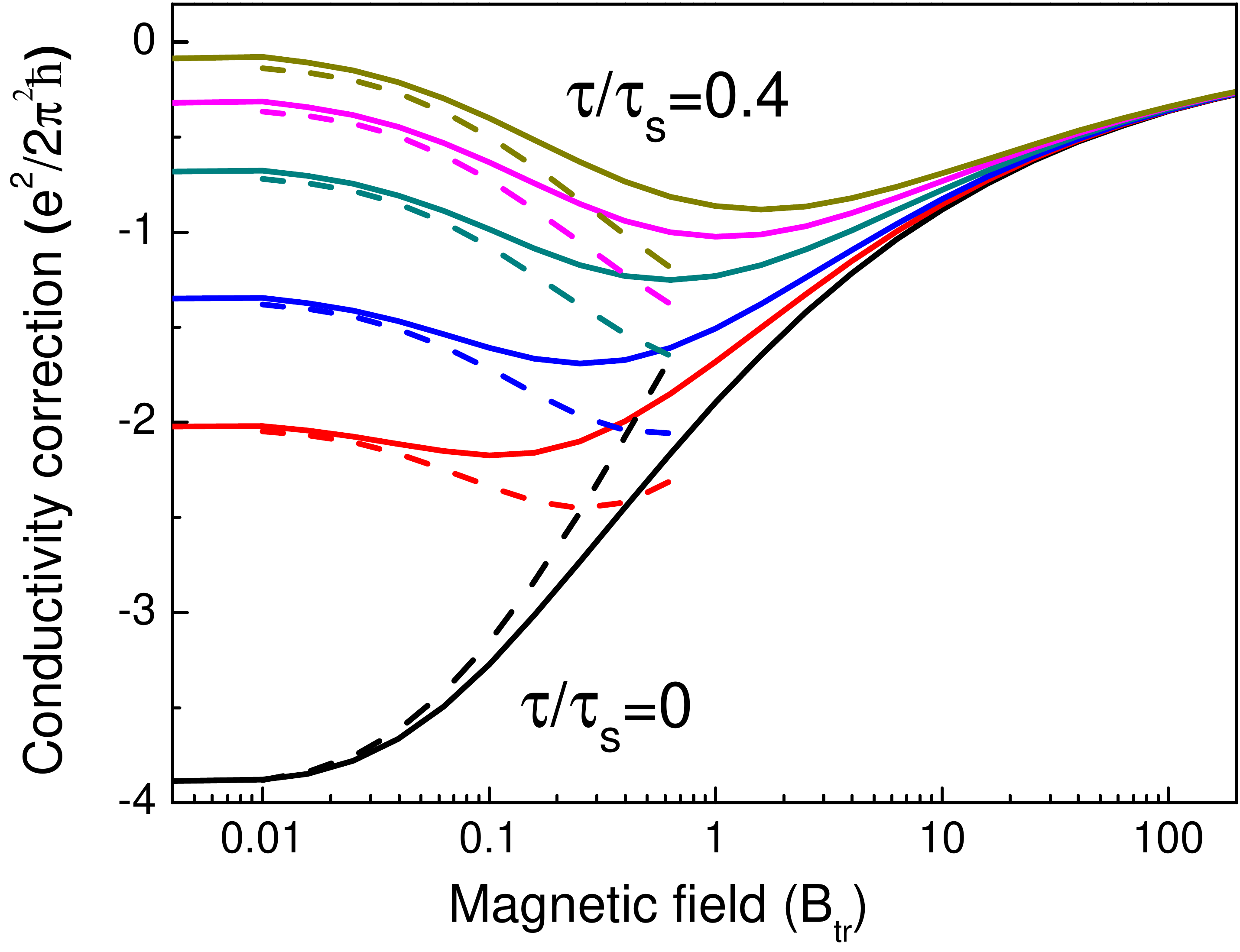}
\caption{Magnetoconductivity at different values of the spin flip rate $\tau/\tau_s=0,0.05,0.1,0.2,0.3$ and~0.4 (from bottom to top). The dephasing rate $\tau/\tau_\phi=0.01$. Dashed curves are calculated in the diffusion approximation, Eq.~\eqref{diff_sf}.}
\label{fig_random_mc}
\end{figure}

The magnetic field dependence of the conductivity correction is plotted in Fig.~\ref{fig_random_mc} by solid lines. The curves have minima at nonzero spin-flip rate with the positions shifted to higher fields with increase of $\tau/\tau_s$. In high magnetic field the correction is independent of the spin-flip rate and has the asymptotics Eq.~\eqref{hf}.

Our calculations show that the account of $b_N$ in the expression for Cooperon Eq.~\eqref{C_sf} changes the magnetoconductivity by less than 1\% for all studied values of $\tau/\tau_s$. Therefore one can use the simplified expression for the Cooperon:
\begin{multline}
\label{C_simple}
		C_{N} = {P_N} \Biggl({2 \over 1 - P_N + {\tau \over 2\tau_s}P_N } \\
	+ {1 \over 1 - P_N + {\tau \over \tau_s}P_N + {\tau \over 4\tau_s} d_N}
		- {1\over 1 - P_N - {\tau \over 4\tau_s}d_N} \Biggr).
\end{multline}

In the diffusion approximation $B \sim B_{tr} \, {\tau \over \tau_\phi} \ll B_{tr}$, 
the asymptotics~\eqref{diff_approx} is valid for $P_N$, and ${Q_N \approx S_{N} \approx 0}$. 
Therefore we obtain the traditional expression for the magnetoconductivity:~\cite{HLN}
\begin{multline}
\label{diff_sf}
	\sigma(B) - \sigma(0) =  {e^2 \over 4 \pi^2 \hbar} 
 \Biggl[f_2\left({B/B_{tr} \over \tau/\tau_\phi + {\tau / \tau_s}} \right) \\ - f_2\left({B/B_{tr} \over \tau/\tau_\phi} \right) + 2f_2\left({B/B_{tr} \over \tau/\tau_\phi + {\tau / 2\tau_s}} \right) \Biggr] ,
\end{multline}
where $f_2(x)$ is defined in Eq.~\eqref{f2}. The calculations in the diffusion approximation are shown in Fig.~\ref{fig_random_mc} by dashed lines. One can see that the diffusion approximation describes the low-field parts of the magnetoconductivity curves but does not adequately reproduce the position of the minimum.

\begin{figure}[h]
\includegraphics[width=0.9\linewidth]{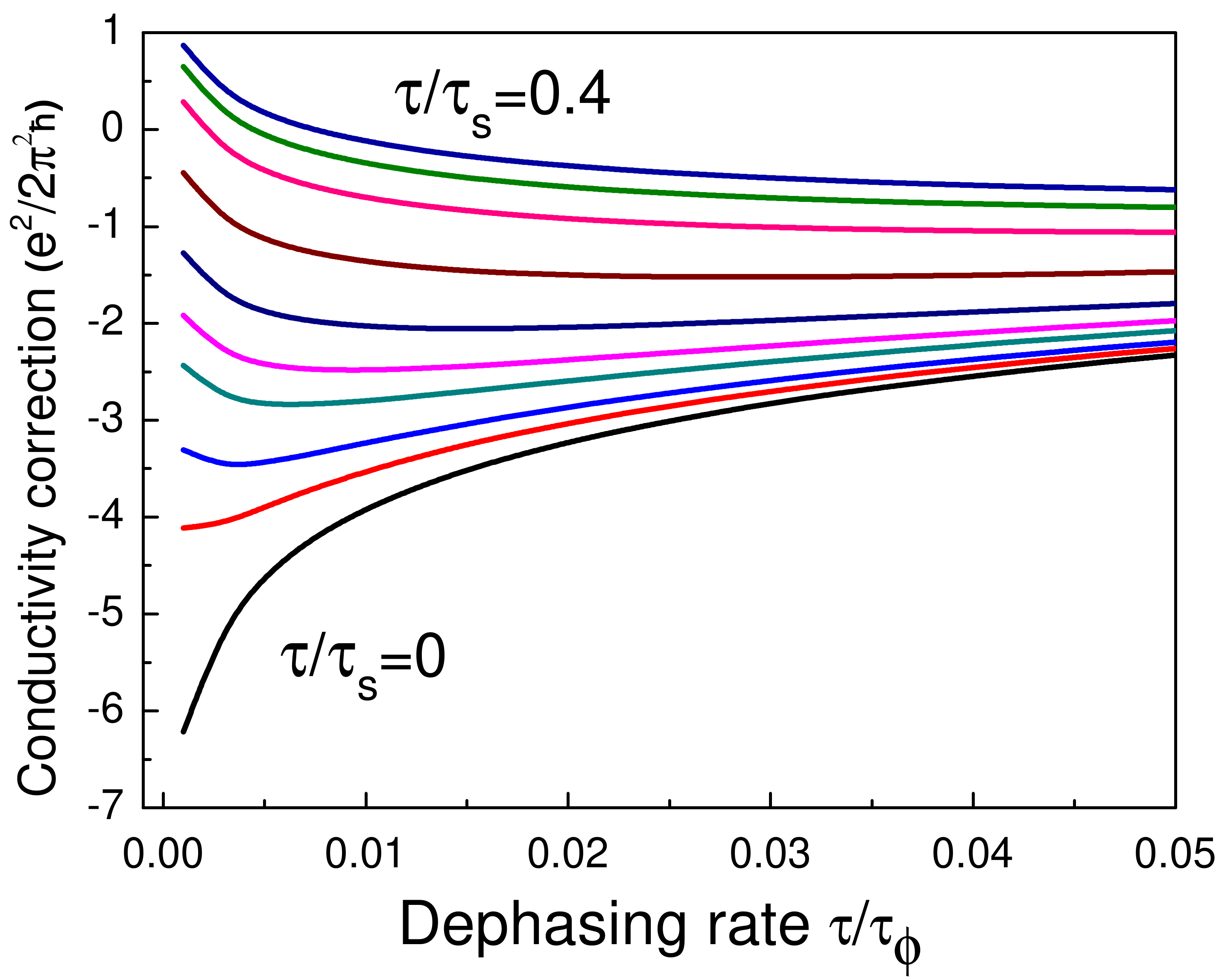}
\caption{Zero-field conductivity correction as a function of the dephasing rate at
the spin flip scattering rate $\tau/\tau_s=0,0.005,0.01,0.02,0.03,0.05,0.1,0.2,0.3$ and~0.4 (from bottom to top).}
\label{fig_zero_field_random}
\end{figure}

In the zero-field limit we have:
\begin{multline}
	\sigma(0) = {e^2 \over 2\pi^2 \hbar} \int\limits_0^{1-\tau/\tau_\phi} dP {P\over 1+P} 
\biggl\{   {1 \over 1 - P + {\tau \over 2\tau_s} Q^2} \\ 
 - {1 \over 1 - P + {\tau \over 2\tau_s}(2P-Q^2)}\\
- \sum_\pm {1 \over 1 - P + {\tau \over 2\tau_s}P \pm {\tau \over 4\tau_s}[S(1+P)+2Q^2]} 
\biggr\}, 
\end{multline}
where $Q^2=P^2(1-P)/(1+P)$ and $S=-Q^2/(2P)$. At $\tau/\tau_s=0$ the correction $\sigma(0)$ is given by Eq.~\eqref{sigma0}. The zero-field correction is plotted in Fig.~\ref{fig_zero_field_random}. 
Since the dephasing rate $1/\tau_\phi$ is proportional to  temperature, Fig.~\ref{fig_zero_field_random} represents the temperature dependence of the weak-localization correction to conductivity.

\section{Linear in wavevector BIA and random SIA splittings}
\label{BIA_linear_random}

It has been demonstrated above that the $\bm k$-linear BIA splitting itself has no effect on the quantum correction to the conductivity. However the $\bm k$-linear BIA spin-orbit interaction~\eqref{H_BIA} may influence the spin dynamics in the presence of spin-flip scattering.
With account for BIA spin splitting, the spin-flip scattering rate Eq.~\eqref{sf_rate} is modified according to~\cite{j_to_s_PRB_11}
\begin{equation}
	{1\over \tau_s(k_0)} = {m \over 2\hbar^3} \left< {\cal K}_{\alpha}(| {\bm k}' - {\bm k} - k_0 \hat{\bm x}|) \left({\bm k}' + {\bm k}\right)^2 \right>_{\varphi_{\bm k},\varphi_{\bm k'}}.
\end{equation}
For short-range spin-orbit disorder, $l_ck_{\rm F} \ll 1$, the correlator ${\cal K}_{\alpha}$ is a constant, and the BIA splitting has no effect on the spin-flip rate. However, for large domains  $l_ck_{\rm F} \gg 1$, the correlator ${\cal K}_{\alpha}(q)=2\pi l_c^2 \left< \alpha^2 \right> \exp(-ql_c)$, and 
it follows from the above expression that
the BIA splitting suppresses spin-flips:~\cite{Poshakinskiy}
\begin{equation}
\label{tau_s_k0}
	{1\over \tau_s(k_0)} ={1\over \tau_s(0)}  \xi[I_0(\xi)K_1(\xi)-K_0(\xi)I_1(\xi)].
\end{equation}
Here $1/\tau_s(0)$ is given by Eq.~\eqref{tau_sf},
$\xi = k_0l_c/2$, and $I_{0,1}$, $K_{0,1}$ are the modified Bessel functions of the first and second kind. At $\xi \geq 1$ the spin-flip time $\tau_s(k_0)$ is longer than $\tau_s(0)$.

Here we examine how the $\bm k$-linear BIA spin splitting changes the anomalous magnetoresistance.
In addition to the above described suppression of the spin-flip rate, the spin-orbit interaction~\eqref{H_BIA} introduces a phase factor into the integrand of Eq.~\eqref{C}: the function $P_0(\bm r_1, \bm r_3)$ is multiplied by 
\[\exp{[{\rm i}S_z k_0 (x_3-x_1)]}.\]
This expression demonstrates that the Cooperons with $S=0$ and with $S=1, S_z=0$ are independent of $\beta$ and, hence, yield the same contribution to the conductivity as in the absence of the BIA splitting.

Two other Cooperons with $S=1, S_z=\pm 1$ depend on $\beta$. We solve the corresponding equations
expanding the Cooperons  by the Landau-level functions  of an electron with the double elementary charge with 
an argument of the oscillator function shifted by $-k_0l_B^2/2$ in the $\bm y$ direction. This allows us to eliminate $\beta$ from one of two coupled Cooperon equations, so it has the form of the first Eq.~\eqref{CCC}. The second Eq.~\eqref{CCC} has the following form at $\beta \neq 0$:
\begin{multline}
\label{eq_Cooperon_Dress}
	\sum_{N_1} \left[ \delta_{NN_1} -  \left(1 - {\tau\over 2\tau_s} \right) P^{(0)}_{NN_1} \right] \overline{C}_{N_1N'}^{(-+)} \\
	= {\tau\over 4\tau_s}  \sum_{N_1} {\rm i}^{N-N_1} P^{(2)}_{NN_1} \overline{C}_{N_1N'}^{(++)} \\
	+ {\tau\over 4\tau_s} \sum_{N_1,N_2} \left( 2P^{(1)}_{NN_2}P^{(1)}_{N_2N_1} + P^{(2)}_{NN_2}P^{(0)}_{N_2N_1} \right)\overline{C}_{N_1N'}^{(++)}.
\end{multline}
Here  ${\cal K}_\alpha$ is assumed to be a constant for brevity, and
\begin{multline}
	P^{(m)}_{NN'} =
\sqrt{{N! \over N'!}} {l_B \over l} \int\limits_0^\infty dx
    \exp{\left( -x {l_B \over \tilde{l} } - {x^2 \over 2}\right)}\\
		\times x^{N'-N} L_N^{N'-N}(x^2) J_{N'-N+m}(2k_0l_B x).
\end{multline}
At $k_0=0$, we have $J_{N'-N+m}(2k_0l_B x) = \delta_{N'+m,N}$, i.e. $P^{(m)}_{NN'}=\Pi_{N,N-m}\delta_{N'+m,N}$, and we come to Eq.~\eqref{CCC} obtained at $\beta=0$.
In the opposite limit $k_0l \gg 1$ we obtain $P^{(m)}_{NN'}=0$ due to rapidly oscillating Bessel functions. As a result we have $\overline{C}_{N+2,N'}^{(-+)} =0$, and $\overline{C}_{NN}^{(++)} = \overline{C}_{NN}^{(--)} =1/a_N$. In this case, the conductivity correction is defined by the Cooperon given by Eq.~\eqref{C_simple}.
For calculation of the conductivity correction at intermediate $k_0l \sim 1$ one should obtain the two Cooperons solving the infinite equation system~\eqref{eq_Cooperon_Dress}.

However, the results of calculation at $k_0=0$ and ${k_0=\infty}$ differ by less than 1~\%. This occurs because $k_0$ is present in the second Cooperon equation only where it is multiplied by the factor $\sim (\tau/\tau_s)b_N$. This factor has almost no effect on the conductivity correction as it is discussed in the previous section, therefore the Cooperon can be taken in the simple form Eq.~\eqref{C_simple} at any value of $k_0l$. 

We see that the $\bm k$-linear BIA splitting has practically no effect on the magnetoconductivity for spin flips at small domains. At large domain size, $l_c k_{\rm F} \gg 1$, the BIA spin splitting influences the magnetoresistance only via renormalization of the spin flip rate $1/\tau_s(k_0)$, Eq.~\eqref{tau_s_k0}, which should be substituted into  Eq.~\eqref{C_simple}.

\section{Discussion}
\label{disc}

In two previous Sections we developed the weak-localization theory for short-range domains with $l_c \ll l$. In the opposite limit $l_c \gg l$, electrons diffusively pass each domain with a nonzero Rashba splitting. 
Therefore, neglecting $\bm k$-cubic terms, the electron energy spectrum has an anisotropic splitting $2\sqrt{(\beta^2+\alpha^2)k_x^2+\alpha^2k_y^2}$, and the conductivity correction inside each domain is described by the  expression $\sigma_{hom}(B;\beta^2+\alpha^2,\alpha^2)$  
where $\sigma_{hom}(B;\beta_+^2,\beta_-^2)$ is the function 
derived 
in Ref.~\onlinecite{MG_LG_FTP} for homogeneous systems with the splitting $2\sqrt{\beta_+^2k_x^2+\beta_-^2k_y^2}$.
If
$l_s \gg l_c \gg l$, 
where $l_s = \hbar^2/(m\left< \alpha^2\right>)$ is the characteristic spin dephasing length with angular brackets denoting  averaging over the spin-orbit disorder,
then the spin rotation angle inside a domain is small. In this limit, the total correction is given by  
$\sigma_{hom}(B;\beta^2,\left< \alpha^2\right>)$. Here we take into account that while the SIA and BIA splitting in some domains can be comparable, the average $\left< \alpha^2\right>$ is much smaller than $\beta^2$.~\cite{Glazov_review} However, account for the random Rashba splitting in the magnetoconductivity is crucial since it gives rise to the change of its sign because $\sigma_{hom}(B;\beta^2,0)$ is definitely positive.~\cite{MG_LG_FTP}
In the opposite limit $l_c \gg l_s$ the interference occurs inside individual domains, and the conductivity correction is given by $\left<\sigma_{hom}(B;\beta^2+\alpha^2,\alpha^2)\right>$. This order of averaging  
is analogous to averaging of the spin polarization in the spin dynamics.~\cite{Poshakinskiy} 

The anomalous magnetoresistance studied here takes place in low magnetic fields $B \sim 1-10$~mT,~\cite{Guzenko_InGaAs,Stud_InGaAs,Koga_InGaAs,HgTe_APL,Spirito_GaN,HgTe_JAP} and it has a sharp field dependence. This allows one to distinguish experimentally between the weak-localization induced magnetoconductivity and the classical one proposed for the same system in Ref.~\onlinecite{class_NMR} which takes place at much higher fields  $B\geq 1$~T. 

Asymmetrically doped (001) QWs with equal BIA and average SIA spin-orbit splittings represent a similar low-symmetry system since the spin-orbit Hamiltonian can be presented in the form $\beta_1 \sigma_z k_x$
where both $x$ and $z$ axes lie in the QW plane. This Hamiltonian is
equivalent to the $\bm k$-linear part of the BIA Hamiltonian~\eqref{H_BIA}, therefore 
the $z$ spin component does not relax by the D'yakonov-Perel mechanism, and 
the weak-localization induced magnetoconductivity is positive.~\cite{LG_05} 
The random spin-orbit interaction~\eqref{H_SO_random} switches on  spin relaxation of the $z$ component with the 
time $2\tau_s(k_0)$ twice longer than one given by Eq.~\eqref{tau_s_k0}. Spin-orbit disorder can be responsible for a small positive part of the magnetoconductivity in the most symmetrical QW with equal BIA and average SIA $\bm k$-linear spin-orbit splittings studied in Ref.~\onlinecite{Koga_InGaAs}.
For developing the weak-localization theory for this case one should solve the Cooperon equations with the correlator obtained from Eq.~\eqref{W} by the substitution $S_x \to S_z$. Therefore the singlet Cooperon has the same form~\eqref{singlet_Cooperon} as in (110) QWs. In contrast, three other Cooperons corresponding to $S=1$ should be found from a coupled equation system~\eqref{C_av} which is infinite in this case.
Besides, the $\bm k$-cubic terms may be also important in (001) QWs, the corresponding theory of weak localization is developed in Ref.~\onlinecite{MG_LG_FTP}.

\section{Conclusions}
\label{concl}

To conclude, theory of weak localization is developed for low symmetrical QWs beyond the diffusion regime.
We demonstrated that the $\bm k$-cubic BIA spin-orbit splitting suppresses the anomalous magnetoresistance in symmetrical (110) QWs and in asymmetrical (111) QWs with zero $\bm k$-linear splitting, but it is not reduced to additional dephasing. The derived expressions are valid for a description of the magnetoconductivity in the whole range of classically weak magnetic fields. We show that the random spin-orbit coupling has a strong effect on weak localization and may result in the positive magnetoresistance even in macroscopically symmetric QWs. The simple expression for  the anomalous magnetoconductivity is derived which is shown to be valid at any value of the spin-flip rate and magnetic field. The effect of $\bm k$-linear splitting on weak localization in the presence of spin flips is found to be 
reduced to renormalization of the spin flip rate without a significant effect of the additional spin-dependent phase.

\acknowledgments 
We thank M.~M. Glazov, E.~Ya.~Sherman M.~O.~Nestoklon for helpful discussions. 
The work was supported by RFBR. 

\end{document}